\documentclass[12pt]{article}

\usepackage{amsmath}
\usepackage{amssymb}
\usepackage{amsthm}
\usepackage{graphicx}
\usepackage{upref}

\newcommand\Cal[1]{{\cal #1}}

\newcommand\reals{\mathbb R}

\newcommand\complex{\mathbb C}

\newcommand\ket[1]{|#1 \rangle}
\newcommand\bra[1]{\langle #1 |}

\newcommand\w{{\rm quant}}

\def\d{{\delta}}

\def\c{{\gamma}}
\def\b{{\beta}}
\def\a{{\alpha}}
\def\e{{\varepsilon}}
\def\l{{\lambda}}

\def\L{{\mathbb L}}
\def\Q{{\bf Q}}

\title{Quantum Algorithms and Complexity\\ 
for Continuous Problems\footnote{To appear in the Springer Encyclopedia
of Complexity and Systems Science}}
\author{A. Papageorgiou\footnote{ap@cs.columbia.edu} 
\ and 
J. F. Traub\footnote{traub@cs.columbia.edu}\\
Department of Computer Science\\
Columbia University\\
New York, USA
}

\date{\ }
\begin{document}

\maketitle



\section*{Article Outline}

\ 
 
Glossary

I. Abstract

II. Introduction

III. Overview of Quantum Algorithms

IV. Integration

V. Path Integration

VI. Feynman-Kac Path Integration

VII. Eigenvalue Approximation

VIII. Qubit Complexity

IX. Approximation

X. Partial Differential Equations

XI. Ordinary Differential Equations

XII. Gradient Estimation

XIII. Simulation of Quantum Systems on Quantum Computers

XIV. Future Directions

Bibliography

\section*{Glossary}
\begin{itemize}
\item {\bf Black Box Model.} 
This model assumes we can collect knowledge about an
input $f$ thru queries without knowing how the answer to the query is computed.
A synonym for black box is oracle.
\item {\bf Classical Computer.} 
A computer which does not use the principles of 
quantum computing to carry out its computations.
\item {\bf Computational Complexity.} In this article, complexity for brevity.
The minimal cost of solving a problem by an algorithm. Some authors use the
word complexity when cost would be preferable. An upper bound on the 
complexity is given by the cost of an algorithm. A lower bound is given by a
theorem which states there cannot be an algorithm which does better.
\item {\bf Continuous Problem.} 
A problem involving real or complex functions of real
or complex variables. Examples of continuous problem are integrals, path
integrals, and partial differential equations.
\item {\bf Cost of an Algorithm.} 
The price of executing an algorithm. The cost 
depends on the model of computation.
\item {\bf Discrete Problem.} A problem whose inputs are from a countable 
set. Examples of discrete problems are integer factorization, traveling
salesman and satisfiability.
\item {\bf $\e$-Approximation.} Most real-world continuous problems can only be
solved numerically and therefore approximately, that is to within an 
error threshold $\e$. The definition of $\e$-approximation depends on
the setting. See worst-case setting, randomized setting, quantum setting.
\item {\bf Information-Based Complexity.} 
The discipline that studies algorithms and
complexity of continuous problems.
\item {\bf Model of Computation.} The rules stating what is permitted in a 
computation and how much it costs. The model
of computation is an abstraction of a physical computer. Examples of models
are Turing machines, real number model, quantum circuit model.
\item {\bf Optimal Algorithm.} 
An algorithm whose cost equals the complexity of the
problem.
\item {\bf Promise.} A statement of what is known about a problem 
a priori before any
queries are made. An example in quantum computation is the promise
that an unknown $1$-bit function is constant or balanced. In 
information-based complexity a promise is also called global information.
\item {\bf Quantum Computing Speedup.} 
The amount by which a quantum computer can 
solve a problem faster than a classical computer. To compute the speedup one 
must know the classical complexity and it's desirable to also know the quantum
complexity.
Grover proved quadratic speedup for search in an unstructured database. Its 
only conjectured that Shor's algorithm provides exponential speedup for integer
factorization since the classical complexity is unknown.
\item {\bf Query.} One obtains knowledge about a particular input
thru queries. For example, if the problem is numerical approximation of
$\int_0^1 f(x)\,dx$ a query might be the evaluation of $f$ at a point. 
In information-based complexity the same concept is called an 
information operation.
\item {\bf Quantum Setting.} There are a number of quantum settings. An example
is a guarantee of error at most $\e$ with probability 
greater than $\tfrac 12$. 
\item {\bf Qubit Complexity.} The minimal number of qubits to solve a problem.
\item {\bf Query Complexity.} The minimal number of queries required to solve 
the problem.
\item {\bf Randomized Setting.} 
In this setting the expected error with respect to 
the probability measure generating the random variables is at most $\e$. The
computation is randomized. An important example of a randomized algorithm is 
the Monte Carlo method.
\item {\bf Worst-Case Setting.} In this setting an error of at most $\e$ is 
guaranteed for all inputs satisfying the promise. The computation is 
deterministic.
\end{itemize}

\section{Abstract}

Most continuous mathematical formulations arising in science and engineering
can only be solved numerically and therefore approximately. We shall always
assume that we're dealing with a numerical approximation to the solution.

There are two major motivations for studying quantum algorithms and
complexity for continuous problems.
\begin{enumerate}
\item Are quantum computers more powerful than classical computers for 
important scientific problems? How much more powerful? 

\item Many important scientific and engineering problems have continuous 
formulations. These problems occur in fields such as physics, chemistry,
engineering and finance.
The continuous formulations include path integration, partial 
differential equations (in particular, the Schr\"odinger equation) and
continuous optimization.
\end{enumerate}

To answer the first question we must know the classical computational 
complexity (for brevity, complexity) of the problem. There have been decades 
of research on the classical complexity of continuous problems in the
field of information-based complexity.
The reason we know the complexity of many 
continuous problems is that we can use adversary arguments to 
obtain their query complexity. 
This may be contrasted with the classical
complexity of discrete problems where we have only conjectures such
as $\mbox{P}\ne\mbox{NP}$. Even the classical complexity of the factorization
of large integers is unknown. Knowing the classical complexity of a
continuous problem we obtain the quantum computation speedup if we know
the quantum complexity. If we know an upper bound on the quantum complexity
through the cost of a particular quantum algorithm then we can obtain a lower
bound on the quantum speedup. 

Regarding the second motivation, 
in this article we'll report on high-dimensional integration, path
integration, Feynman path integration, the smallest eigenvalue of a
differential equation, approximation, partial differential equations,
ordinary differential equations and gradient estimation. 
We'll also briefly report
on the simulation of quantum systems on a quantum computer.

\section{Introduction}

We provide a summary of the contents of the article.
\vskip 0.5pc

\noindent 
Section~\ref{sec:Overview}: Overview of quantum algorithms and complexity.

We define basic concepts and notation including quantum algorithm, continuous
problem, query complexity and qubit complexity.
\vskip 0.5pc

\noindent Section~\ref{sec:Integration}: Integration

High-dimensional integration, often in hundreds or thousands of variables, 
is one of the most commonly occurring continuous problems
in science. In Section~\ref{sec:IntClass} we report on complexity results
on a classical computer. For illustration we begin with a one-dimensional 
problem and give a simple example of an adversary argument. We then move on
the $d$-dimensional case and indicate that in the worst case the complexity
is exponential in $d$; the problem suffers the curse of dimensionality.
The curse can be broken by the Monte Carlo algorithm. In 
Section~\ref{sec:IntQuant} we report on the algorithms and complexity results
on a quantum computer. Under certain assumptions on the
promise the quantum query complexity enjoys exponential speedup over 
classical worst case query complexity.

A number of the problems we'll discuss enjoy exponential quantum query speedup.
This does not contradict Beals et al. \cite{beals} who prove that speedup
can only be polynomial. The reason is that \cite{beals} deals with
problems concerning total Boolean functions.

For many classes of integrands there is quadratic speedup 
over the classical randomized query complexity. 
This is the same speedup as 
enjoyed by Grover's search algorithm of an unstructured database. To obtain the
quantum query complexity one has to give matching upper and lower bounds. As
usual the upper bound is given by an algorithm, the lower bound by a
theorem. The upper bound is given by the amplitude amplification algorithm of 
Brassard et al. \cite{BHMT}. We outline a method
for approximating high-dimensional integrals using this algorithm. The
quantum query complexity lower bounds for integration are based on the lower 
bounds of Nayak and Wu \cite{nayak} for computing the mean of a 
Boolean function.
\vskip 0.5pc

\noindent Section~\ref{sec:PathInt}: Path Integration

We define a path integral and provide an example due to Feynman. In 
Section~\ref{sec:PIClass} we report on complexity results on
a classical computer. If the promise is that the class of integrands has finite
smoothness, then  path integration is intractable in the worst case. 
If the promise is 
that the class of integrands consists of entire functions the query complexity
is tractable even in the worst case. For smooth functions intractability
is broken by Monte Carlo. In Section~\ref{sec:PIQuant} we report on the
algorithm and complexity on a quantum computer. The quantum query complexity 
enjoys Grover-type speedup over classical randomized query complexity. We 
outline the quantum algorithm.
\vskip 0.5pc

\noindent Section~\ref{sec:Feynman}: Feynman-Kac path integration.

The Feynman-Kac path integral provides the solution to the diffusion equation.
In Section~\ref{sec:FKClass} we report on algorithms and complexity on a
classical computer. In the worst case for a $d$-dimensional Feynman-Kac path 
integral the problem again suffers the curse of dimensionality which can be
broken by Monte Carlo. In Section~\ref{sec:FKQuant} we indicate the algorithm 
and query complexity on a quantum computer.
\vskip 0.5pc

\noindent Section~\ref{sec:Eig}: Eigenvalue approximation.

One of the most important problems in physics and chemistry is approximating
the ground state energy governed by a differential equation. Typically, 
numerical algorithms on a classical computer are well known. Our focus is on 
the Sturm-Liouville eigenvalue (SLE) problem where the first complexity
results were recently obtained. The SLE equation is also called the 
time-independent Schr\"odinger equation. In
Section~\ref{sec:EigClass} we present an algorithm on a classical computer.
The worst case query complexity suffers the curse of dimensionality. We also
state a randomized algorithm. The randomized query complexity is unknown
for $d>2$
and is an important open question. In Section~\ref{sec:EigQuant} we
outline an algorithm for a quantum computer. The quantum query complexity is 
not known when $d>1$. 
It has been shown that it is not exponential in $d$; the problem is tractable 
on a quantum computer.
\vskip 0.5pc

\noindent Section~\ref{sec:Qubit}: Qubit complexity.

So far we've focused on query complexity. For the foreseeable future the
number of qubits will be a crucial computational resource. We give a
general lower bound on the qubit complexity of continuous problems. We
show that because of this lower bound there's a problem that cannot be
solved on a quantum computer but that's easy to solve on a classical
computer using Monte Carlo. 

A definition of a quantum algorithm is given by (\ref{eq:qa1}); the queries 
are deterministic. 
Wo\'zniakowski \cite{W06} introduced the quantum setting with randomized
queries for continuous problems. 
For path integration there is an exponential reduction in the
qubit complexity.
\vskip 0.5pc

\noindent Section~\ref{sec:Approx}: Approximation.

Approximating functions of $d$ variables is a fundamental and generally hard 
problem. The complexity is sensitive to the norm, $p$, on the class
of functions and to several other parameters.
For example, if $p=\infty$ approximation suffers the
curse of dimensionality in the worst and randomized classical cases. The 
problem remains intractable in the quantum setting.
\vskip 0.5pc

\noindent Section~\ref{sec:PDE}: Partial Differential Equations.

Elliptic partial differential equations have many important applications and
have been extensively studied. In particular, 
consider an equation of order $2m$
in $d$ variables. In the classical worst case setting the problem is 
intractable. The classical randomized and quantum settings were only
recently studied. The conclusion is that the quantum may or may not provide a 
polynomial speedup; it depends on certain parameters.
\vskip 0.5pc

\noindent Section~\ref{sec:IVP}: Ordinary Differential Equations.

Consider a system of initial value ordinary equations in $d$ variables. 
Assume that the right hand sides satisfy a H\"older condition. The problem
is tractable even in the worst case with the exponent of $\e^{-1}$ depending
on the H\"older class parameters. The complexity of classical randomized and
quantum algorithms have only recently been obtained. The quantum setting yields
a polynomial speedup.
\vskip 0.5pc

\noindent Section~\ref{sec:Gradient}: Gradient estimation.

Jordan \cite{Jo} showed that approximating the gradient of a function can be
done with a single query on a quantum computer although it takes $d+1$
function evaluations on a classical computer. A simplified version of Jordan's algorithm is presented.
\vskip 0.5pc

\noindent Section~\ref{sec:Sim}: Simulation of quantum systems on quantum 
computers.

There is a large and varied literature on simulation of quantum systems on 
quantum computers. The focus in these papers is typically on the cost
of particular classical and quantum algorithms without complexity analysis
and therefore without speedup results. To give the reader a taste of this 
area we list some sample papers.
\vskip 0.5pc

\noindent Section~\ref{sec:FutDir}: Future directions.

We briefly indicate a number of open questions.

\section{Overview of Quantum Algorithms}\label{sec:Overview} 

A quantum algorithm consists of a sequence of unitary
transformations applied to an initial state. The result of
the algorithm is obtained by measuring its final state.
The quantum model of computation is discussed in detail in
\cite{beals,bennet,bernstein,cleve,heinrich,NC} and we
summarize it here as it applies to continuous problems.

The initial state $|\psi_0\rangle$ of the algorithm is a unit vector 
of the Hilbert space
$\Cal{H}_\nu=\complex^2\otimes \cdots\otimes \complex^2$, $\nu$ times,  
for some appropriately chosen integer $\nu$, where $\complex^2$ is the 
two dimensional space of complex numbers. The dimension of
$\Cal{H}_\nu$ is $2^{\nu}$. The number $\nu$ denotes the number of 
qubits used by the quantum algorithm. 

The final state $|\psi\rangle$ is also a unit vector of
$\Cal{H}_\nu$ and is obtained from the initial state 
$|\psi_0\rangle$ through
a sequence of unitary $2^{\nu}\times 2^{\nu}$ matrices, i.e.,
\begin{equation}
|\psi\rangle_f\,:=\,U_TQ_fU_{T-1}Q_f\cdots U_1Q_fU_0 |\psi_0\rangle.
\label{eq:qa1}
\end{equation}
The unitary matrix $Q_f$  
is called a quantum query and is used to provide information to the
algorithm about a function $f$. 
$Q_f$ depends on $n$ function evaluations $f(t_1),\dots,f(t_n)$,
at deterministically chosen points, $n\le 2^{\nu}$. 
The $U_0,U_1,\dots,U_T$ are unitary matrices that do not depend 
on $f$. 
The integer $T$ denotes the number of quantum
queries.

For algorithms solving discrete problems, 
such as Grover's algorithm for the search of an unordered database
\cite{Grover}, the input $f$ is considered to be a Boolean function.
The most commonly studied quantum query is the {\it bit} query.
For a Boolean function $f:\{0,\dots,2^m-1\}\to\{0,1\}$, 
the bit query is defined by  
\begin{equation}
Q_f|j\rangle|k\rangle\,=\,|j\rangle|k\oplus f(j)\rangle. 
\label{eq:boolquery}
\end{equation}
Here $\nu=m+1$, $|j\rangle\in\Cal{H}_m$, and
$|k\rangle\in\Cal{H}_{1}$ with $\oplus$ denoting addition
modulo $2$. For a real function $f$ the query
is constructed by taking the 
most significant bits of the function $f$ evaluated at some points
$t_j$.
More precisely, as in \cite{heinrich}, the bit query for $f$ has the
form
\begin{equation}
Q_f|j\rangle|k\rangle\,=\,|j\rangle|k\oplus \beta(f(\tau(j)))\rangle, 
\label{eq:bitquerydef}
\end{equation}
where the number of qubits is now $\nu=m'+m''$ and $|j\rangle\in
\Cal{H}_{m'}$, $|k\rangle\in\Cal{H}_{m''}$. The functions
$\beta$ and $\tau$ are used to discretize the domain $\mathcal{D}$ 
and the range $\mathcal{R}$ of $f$, respectively.
Therefore,
$\beta:\mathcal{R}\to\{0,1,\dots,2^{m''}-1\}$ and
$\tau:\{0,1,\dots,2^{m'}-1\}\to\mathcal{D}$.
Hence, we compute $f$ at 
$t_j=\tau(j)$ and then take the $m''$ most significant bits of
$f(t_j)$ by $\beta(f(t_j))$, for the details and the possible use
of ancillary qubits see  \cite{heinrich}.

At the end of the quantum algorithm, 
the final state $|\psi_f \rangle$ is measured.
The measurement produces one of $M$ outcomes, where $M\le 2^{\nu}$. 
Outcome $j\in\{0,1,\dots,M-1\}$ occurs with
probability $p_f(j)$, which depends on $j$ and the input $f$.
Knowing the outcome $j$, we classically compute
the final result $\phi_f(j)$ of the algorithm. 

In principle, quantum algorithms may have measurements
applied between sequences of unitary transformations of the form 
presented above. However, any algorithm with multiple measurements 
can be simulated by
a quantum algorithm with only one measurement \cite{bernstein}.

Let $S$ be a linear or nonlinear operator such that
\begin{equation}
S:\cal{F}\to \cal{G}.
\label{eq:contprob}
\end{equation}
Typically, $\cal{F}$ is a linear space of real functions 
of several variables,
and $\cal{G}$ is a normed linear space.
We wish to approximate $S(f)$ to within $\e$ for $f\in\cal{F}$.
We approximate $S(f)$ using
$n$ function evaluations $f(t_1),\dots,f(t_n)$ at deterministically
and a priori
chosen sample points. The quantum query $Q_f$ encodes this information, 
and the quantum algorithm obtains this information from $Q_f$.

Without
loss of generality, we consider algorithms that approximate $S(f)$ with
probability $p\ge\tfrac34$. 
We can boost the success probability of an algorithm to become 
arbitrarily close to one
by repeating the algorithm a number of times. The success probability becomes
at least $1-\d$ with a number of repetitions proportional to $\log\d^{-1}$.

The local error of the quantum algorithm (\ref{eq:qa1})
that computes the approximation $\phi_f(j)$, for
$f\in \cal{F}$ and the outcome $j\in\{0,1,\dots,M-1\}$, is defined by 
\begin{equation}
e(\phi_f,S)\,=\,\min \bigg\{\, \a :\quad \sum_{j:\ \|S(f) - 
\phi_f(j)\| \,\le\, \a\,}p_f(j)\geq \tfrac34\,\bigg\},
\label{eq:localperr}
\end{equation}
where $p_f(j)$ denotes the probability of obtaining outcome $j$ 
for the function $f$.
The worst case error of a quantum algorithm $\phi$ 
is defined by
\begin{equation}
e^{\w}(\phi,S)\,=\,\sup_{f\in \cal{F}}
e(\phi_f,S).
\label{eq:wperr}
\end{equation}
The query complexity ${\rm comp}^{\rm query}(\e,S)$ 
of the problem $S$ is the minimal number of queries necessary for 
approximating the solution with accuracy
$\e$, i.e.,
\begin{equation}
{\rm comp}^{\rm query}(\e)\,=\,\min\{\,T:\ \exists\ \phi \ \mbox{such that}\
e^{\w}(\phi,S)\,\le\,\e\;\}.
\label{eq:querycomp}
\end{equation}
Similarly, the qubit complexity of the problem $S$ is the minimal number
of qubits necessary for approximating the solution with accuracy
$\e$, i.e.,
\begin{equation}
{\rm comp}^{\rm qubit}(\e)\,=\,\min\{\,\nu :\ \exists\ \phi \ 
\mbox{such that}\
e^{\w}(\phi,S)\,\le\,\e\;\}.
\label{eq:qubitcomp}
\end{equation}

\section{Integration}\label{sec:Integration}

Integration is one of the most commonly occurring mathematical problems. 
One reason is that
when one seeks the expectation of a continuous process one has to compute an 
integral. Often the integrals are over hundreds or thousands of variables.
Path integrals are taken over an infinite number of variables. See 
Section~\ref{sec:PathInt}.

\subsection{Classical Computer}\label{sec:IntClass}

We begin with a one dimensional example to illustrate some basic concepts
before moving to the $d$-dimensional case (Number of dimensions and number of 
variables are used interchangeably.) Our simple example is to approximate
$$I(f)=\int_0^1 f(x)\, dx.$$
For most integrands we can't use the fundamental theorem of calculus to
compute the integral analytically; we have to approximate numerically
(most real world continuous problems have to be approximated numerically). 
We have to make a promise about $f$. Assume 
$$F_1=\left\{ f:[0,1]\to\reals\; |\;  \mbox{continuous and}\; 
|f(x)|\le 1, \; x\in [0,1] \right\}.$$
Use  queries to compute
$$y_f=[f(x_1),\dots,f(x_n)].$$
We show that with this promise one cannot guarantee an $\e$-approximation on a
classical computer. We use a simple adversary argument. Choose arbitrary 
numbers $x_1,\dots,x_n\in [0,1]$. The adversary answers all these queries by
answering $f(x_i)=0$, $i=1,\dots,n$.

\begin{figure}[!h]
\centering
\includegraphics[width=2.75in, height=2.5in]{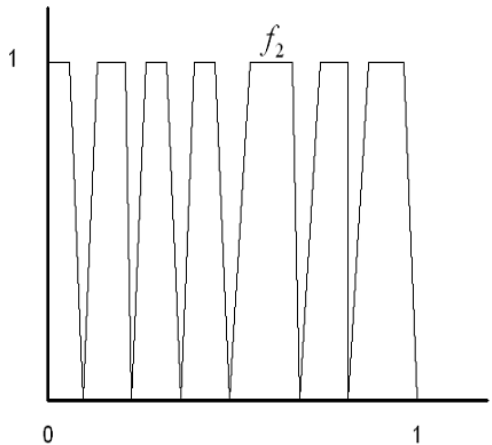}
\caption{All the function evaluations are equal to zero but the integral is 
close to one}
\label{fig:f1}
\end{figure}

What is $f$? It could be $f_1\equiv 0$ and $\int_0^1 f_1(x)\,dx
=0$ or it could be the $f_2$ shown in Figure~\ref{fig:f1}. 
The value of $\int_0^1f_2(x)\, dx$ can be arbitrarily close to $1$. 
Since $f_1(x)$ and $f_2(x)$
are indistinguishable with these query answers and this promise, it is 
impossible to guarantee an $\e$-approximation on a classical computer
with $\e < \tfrac 12$.
We will return to this example in Section~\ref{sec:Qubit}.

We move on to the $d$-dimensional case. Assume that the integration limits are
finite. For simplicity we  assume we're integrating
over the unit cube $[0,1]^d$. So our problem is
$$I(f)=\int_{[0,1]^d}f(x)\, dx.$$
If our promise is that $f\in F_d$, where
$$F_d=\left\{ f:[0,1]^d\to\reals\; |\;  \mbox{continuous and}\; 
|f(x)|\le 1, \; x\in [0,1]^d \right\},$$
then we can't 
compute an $\e$-approximation regardless of the value of $d$. 
Our promise has to be stronger. Assume that 
integrand class has {\it smoothness} $r$. There are various ways to define
smoothness $r$ for functions of $d$ variables. See, for example, the 
definition 
\cite[p. 25]{TW98}. (For other definitions see \cite{TWW88}.) 
With this definition Bakhvalov \cite{Bakh77} showed that the query 
complexity is 
\begin{equation}
{\rm comp}^{\rm query}_{\rm clas-wor}(\e)= \Theta\left(\e^{-d/r}\right).
\end{equation}
This is the worst case query complexity on a classical computer.

What does this formula imply? If $r=0$ the promise is that the functions are
only continuous but have no smoothness. 
Then the query complexity is $\infty$, that
is, we cannot guarantee an $\e$-approximation. If $r$ and $\e$ are fixed
the query complexity is an exponential function of~$d$. We say the problem 
is intractable. Following R. Bellman this is also called the {\it curse of
dimensionality}. In particular, let $r=1$. Then the promise is that the 
integrands have one derivative and the query complexity is $\Theta(\e^{-d})$.

The curse of dimensionality is present for many for continuous problems 
in the worst 
case  setting. Breaking the curse is one of the central issues of 
information-based complexity. For high-dimensional integration the curse can 
be broken for $F_d$ by the Monte Carlo algorithm which is a special 
case of a randomized algorithm.
The Monte Carlo algorithm is defined by
\begin{equation}
\phi^{\rm MC}(f)=\frac 1n\sum_{i=1}^n f(x_i),
\end{equation}
where the $x_i$ are chosen independently from a uniform distribution. Then 
the expected error is
$$e^{\rm MC}(f)=\frac{\sqrt{{\rm var}(f)}}{\sqrt{n}},$$
where
$${\rm var}(f)=\int_{[0,1]^d}f^2(x)\, dx -\left\{
\int_{[0,1]^d}f(x)\,dx\right\}^2$$
is the variance of $f$. For the promise $f\in F_d$
the query complexity is given by
\begin{equation}
{\rm comp}^{\rm query}_{\rm clas-ran}(\e)=\Theta(\e^{-2}).
\label{eq:rancomp}
\end{equation}
This is the randomized query complexity on a classical computer.

Thus Monte Carlo breaks the curse of dimensionality for the integration 
problem;
the problem is tractable. Why should picking sample points at random be much
better than picking them deterministically in the optimal way?
This is possible because we've replaced the guarantee of the worst case
setting by the stochastic assurance of the randomized setting. There's no
free lunch!

The reader will note that (\ref{eq:rancomp}) is a complexity result even 
though it is the cost of a particular algorithm, the Monte Carlo algorithm. 
The reason is that for integrands satisfying this promise Monte Carlo has
been proven optimal. It is known that if the integrands are smoother Monte 
Carlo is not optimal; See \cite[p. 32]{TW98}.

In the worst case setting (deterministic) the query complexity is infinite 
if $f\in F_d$. In the randomized setting the query complexity
is independent of $d$ if $f\in F_d$.
This will provide us guidance when we introduce 
Monte Carlo sampling into the  model of quantum computation in 
Section~\ref{sec:Qubit}.

Generally pseudo-random numbers are used in the implementation of Monte Carlo 
on a classical computer. The quality of a pseudo-random number generator is 
usually tested statistically; see, for example, Knuth \cite{Knuth}. Will
(\ref{eq:rancomp}) still hold if a pseudo-random number generator is
used? An affirmative answer is given by \cite{TW92} provided some care 
is taken in the choice of the generator and $f$ is Lipschitz.

\subsection{Quantum Computer}\label{sec:IntQuant}

We've seen that Monte Carlo breaks the curse of dimensionality for 
high-dimensional integration on a classical computer and that the query 
complexity is of order $\e^{-2}$. Can we do better on a quantum computer?

The short answer is yes. Under certain assumptions on the promises, which will
be made precise below, the quantum query complexity enjoys exponential 
speedup over the classical worst case query complexity and quadratic 
speedup over the classical randomized query complexity. The latter is the
same speedup as enjoyed by Grover's search algorithm of an unstructured
database \cite{Grover}.

To show that the quantum query complexity is of order $\e^{-1}$ we have to give
matching, or close to matching upper and lower bounds. Usually, the upper 
bound is given by an algorithm, the lower bound by a theorem. The upper bound 
is given by the amplitude amplification algorithm of Brassard et al. 
\cite{BHMT} which we describe briefly.

The amplitude amplification algorithm of Brassard et al. computes the 
mean 
$$\mbox{SUM}(f)=\frac 1N \sum_{i=0}^{N-1}f(i),$$
of a Boolean function $f:\{0,1,\dots,N-1\}\to \{0,1\}$, where
$N=2^k$, with error $\e$ and probability at least $8/\pi^2$, 
using a number of queries proportional to $\min\{\e^{-1},N\}$.
Moreover, Nayak and Wu \cite{nayak} show that the order of magnitude of the 
number of queries of this algorithm is optimal.
Without loss of generality we can assume that $\e^{-1} \ll N$.

Perhaps the easiest way to understand the algorithm is
to consider the operator 
$$G=(I - \ket{\psi}\bra{\psi})O_f,$$
that is used in Grover's search algorithm. Here 
$$O_f\ket{x}=(-1)^{f(x)}\ket{x}, \quad x\in\{0,1\}^k,$$
denotes the query, which is slightly different 
yet equivalent \cite{NC} to the one we defined in 
(\ref{eq:boolquery}). Let
$$\ket{\psi}= \sqrt{\frac 1N} \sum_{x}\ket{x}$$
be the equally weighted superposition of all the states.

If $M$ denotes the number of assignments 
for which $f$ has the value $1$ then
$$\mbox{SUM}(f)=\frac MN.$$
Without loss of generality $1\le M\le N-1$.
Consider the space $\Cal H$ spanned by the states
\begin{equation*}
\ket{\psi_0}= \sqrt{\frac 1{N-M}} \sum_{x:f(x)=0}\ket{x}\quad\mbox{and}\quad
\ket{\psi_1}= \sqrt{\frac 1M} \sum_{x:f(x)=1}\ket{x}.
\end{equation*} 
Then
$$\ket{\psi}= \cos(\theta/2) \ket{\psi_0} + \sin(\theta/2)\ket{\psi_1}$$
where $\sin(\theta/2)=\sqrt{M/N}$, and $\theta/2$ is the angle between
the states $\ket{\psi}$ and $\ket{\psi_0}$. 
Thus
$$\sin^2\left( \frac{\theta}2 \right)= \frac MN.$$

Now consider the operator $G$ restricted to $\Cal H$ which has the matrix
representation
\begin{equation*}
G=\left( \begin{array}{cc}
\cos \theta & -\sin \theta\\
\sin \theta  & \cos \theta
\end{array} \right).
\end{equation*}
Its eigenvalues are $\lambda_{\pm}=e^{\pm i\theta}$, $i=\sqrt{-1}$, 
and let $\ket{\xi_{\pm}}$
denote the corresponding eigenvectors. 

We can express $\ket{\psi}$ using the $\ket{\xi_{\pm}}$ to get
$$\ket{\psi}=a \ket{\xi_{-}} + b \ket{\xi_{+}},$$
with $a,b\in\complex$, $|a|^2+|b|^2=1$. 
This implies that phase estimation \cite{NC} with
$G^p$, $p=2^j$, $j=0,\dots,t-1$, and initial state 
$\ket{0}^{\otimes t}\ket{\psi}$ can be 
used to approximate either $\theta$ or $2\pi-\theta$ 
with error proportional to $2^{-t}$. Note that the 
first $t$ qubits of the initial state determine the accuracy of 
phase estimation. 
Indeed, let $\tilde \phi$ be the result of phase estimation. 
Since $\sin^2(\theta/2)= \sin^2(\pi -\theta /2)$,  
$$\left| \sin^2( \pi\tilde \phi) - \frac NM\right| = 
O(2^{-t}),$$
with probability at least $8/\pi^2$; see \cite{BHMT,NC} for the details.
Setting $t=\Theta(\log\e^{-1})$ and observing that phase estimation
requires a number of applications of $G$ (or queries $O_f$) proportional 
to $\e^{-1}$ yields the result. (The complexity of quantum algorithms for the
average case approximation of the Boolean mean has also been studied
\cite{HKW,P04}.)

For a real function $f:\{0,1,\dots,N-1\}\to [0,1],$ we can approximate
$$\mbox{SUM}(f)=\frac 1N \sum_{i=0}^{N-1}f(i),$$
by reducing the problem to the computation of the Boolean mean. 
One way to derive this reduction is
to truncate $f(i)$ to the desired number of significant bits,
typically, polynomial in $\log\e^{-1}$, and
then to use the bit representation of the function values to derive
a Boolean function whose mean is equal to the mean of the truncated
real function, see, e.g. \cite{W06}. The truncation of the function values
is formally expressed through the mapping $\b$ in (\ref{eq:bitquerydef}).
Variations of this idea have been used in the literature 
\cite{AW99,heinrich,N01}.

Similarly, one discretizes the domain of a function $f:[0,1]^d\to [0,1]$
using the function~$\tau$ in (\ref{eq:bitquerydef}) and then uses the
amplitude amplification algorithm to compute the average
\begin{equation}
\mbox{SUM}(f)=\frac 1N \sum_{i=0}^{N-1}f(x_i),
\label{eq:sum}
\end{equation}
for $x_i\in [0,1]^d$, $i=0,\dots,N-1$.

The quantum query complexity of computing the average (\ref{eq:sum})
 is of order $\e^{-1}$. 
On the other hand, the classical deterministic worst case query complexity 
is proportional to $N$ (recall that $\e^{-1}\ll N$), 
and the classical randomized query complexity is proportional to $\e^{-2}$. 

We now turn to the approximation of high-dimensional integrals
and outline an algorithm for solving this problem. 
Suppose $f:[0,1]^d\to [0,1]$ is a function for which we are given some 
promise, for instance, that $f$ has smoothness $r$. The algorithm
integrating $f$ with accuracy $\e$ has two parts. First, using 
function evaluations $f(x_i)$, $i=1,\dots,n$, at deterministic points,
it approximates $f$ classically, by a function $\hat f$ with error $\e_1$, 
i.e., 
$$\| f- \hat f\|\le \e_1,$$
where $\|\cdot\|$ is the $L_\infty$ norm.
The complexity of this problem has been extensively 
studied and there are numerous results \cite{novak,TWW88,TW98} specifying
the optimal choice of $n$ and the points $x_i$ that must be used to
achieve error $\e_1$.
Thus 
$$\int_{[0,1]^d} f(x)\, dx = \int_{[0,1]^d}\hat f(x)\, dx 
+ \int_{[0,1]^d} g(x)\, dx,$$
where $g=f-\hat f$.
Since $\hat f$ is known and depends linearly on the $f(x_i)$ the algorithm
proceeds to integrate it 
exactly. So it suffices to approximate $ \int_{[0,1]^d} g(x)\, dx$ 
knowing that $\|g\|\le \e_1$.

The second part of the algorithm approximates the integral of $g$
using the amplitude amplification algorithm 
to compute
$$\mbox{SUM}(g) = \frac 1N \sum_{i=0}^{N-1} g(y_i),$$
for certain points $y_i\in [0,1]^d$, with error $\e_2$. Once more, 
there are many results, see \cite{TWW88,TW98} for surveys 
specifying the optimal $N$ and the points $y_i$,
so that $\mbox{SUM}(g)$ approximates $\int_{[0,1]^d} g(x)\,dx$ with error
$\e_2$. Finally, the algorithm approximates the original integral 
by the sum of the results of its two parts. 
The overall error of the algorithm is proportional to $\e=\e_1+\e_2$. 

Variations of the
integration algorithm we described are known to have optimal query complexity
for a number of different promises
\cite{heinrich,H03,HN02,N01}. The quantum query complexity lower bounds for integration
are based on the lower bounds of Nayak and Wu \cite{nayak} for 
computing the Boolean mean.
The quantum algorithms offer an exponential speedup over
classical deterministic algorithms and a polynomial speedup over classical
randomized algorithms for the query complexity of 
high-dimensional integration. The table below summarizes the query complexity
(up to polylog factors) of high-dimensional integration 
in the worst case, randomized and quantum 
setting
for functions belonging to H\"older classes $F^{r,\a}_d$ and Sobolev spaces
$W^{r}_{p,d}$.
Heinrich obtained most of the quantum query complexity results in a 
series of papers, which we cited earlier. Heinrich summarized his results in 
\cite{H03b} where a corresponding table showing error bounds can be found.
\begin{center}
\begin{tabular}{|l|l|l|l|}
\hline
             & Worst case          & Randomized             & Quantum \\
\hline 
\hline
$F^{r,\a}_d$ & $\e^{-d/(r+\a)}$    & $\e^{-2d/(2(r+\a)+d)}$ & $\e^{-d/(r+\a+d)}$\\
$W^{r}_{p,d},\quad 2\le p\le\infty$ & $ \e^{-d/r}$ & $\e^{-2d/(2r+d)}$ & $\e^{-d/(r+d)}$ \\
$W^{r}_{p,d},\quad 1 \le p\le 2$ &  $ \e^{-d/r}$ & $\e^{-pd/(rp+pd-d)}$ & $\e^{-d/(r+d)}$ \\
$W^{r}_{1,d}$ & $ \e^{-d/r}$ & $ \e^{-d/r}$ & $\e^{-d/(r+d)}$ \\
\hline
\end{tabular}
\end{center}

Abrams and Williams~\cite{AW99} were the first 
to apply the amplitude amplification algorithm to  
high-dimensional integration.
Novak \cite{N01} was the first to spell out his promises and thus obtained 
the first complexity results for high-dimensional integration.

\section{Path Integration}\label{sec:PathInt}

\begin{figure}[!h]
\centering
\includegraphics[width=2in, height=1.73in]{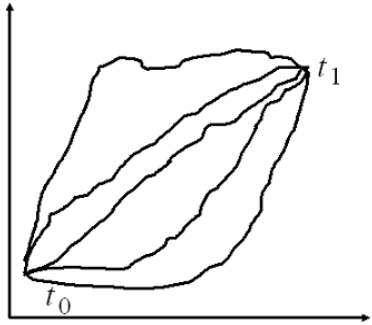}
\caption{Different trajectories of a particle}
\label{fig:f2}
\end{figure}

A path integral is defined as
\begin{equation}
I(f)=\int_X f(x)\, \mu(dx),
\label{eq:pathint}
\end{equation}
where $\mu$ is a probability measure on an infinite-dimensional space $X$. It
can be viewed as an infinite-dimensional integral. For illustration we give an 
example due to R. Feynman. In classical mechanics a particle 
at a certain position
at time $t_0$
has a unique trajectory to its position at time $t_1$. Quantum mechanically 
there are an infinite number of possible trajectories which Feynman called 
histories, see Figure~\ref{fig:f2}. Feynman summed over the histories. 
If one goes to the limit one gets a 
path integral. Setting $t_0=0$, $t_1=1$ this integral is
$$I(f)=\int_{C[0,1]}f(x)\, \mu(dx),$$
which is a special case of (\ref{eq:pathint}). 

Path integration occurs in numerous applications including quantum mechanics, 
quantum chemistry, statistical mechanics, and mathematical finance. 

\subsection{Classical Computer}\label{sec:PIClass}

The first complexity analysis of path integration is due to 
Wasilkowski and Wo\'zniakowski~\cite{WW96}; see 
also \cite[Ch. 5]{TW98}. They studied the deterministic worst case and 
randomized settings and assume that $\mu$ is a Gaussian measure; 
an important special 
case of a Gaussian measure is a Wiener measure. They make the promise that
$F$, the class of integrands, consists of functions $f:X\to\reals$ whose 
$r$-th 
Fr\'echet derivatives are continuous and uniformly bounded by unity.
If $r$ is finite then path integration is intractable. Curbera~\cite{Cur00} 
showed that the worst case query complexity is of order $\e^{\e^{-\b}}$
where $\b$ is a positive number depending on $r$. 

Wasilkowski and Wo\'zniakowski \cite{WW96} also considered the 
promise that $F$ consists of
entire functions and that $\mu$ is the Wiener measure. Then the query 
complexity is a polynomial in~$\e^{-1}$. More precisely, they provide an
algorithm for calculating a worst case $\e$-approximation with cost of 
order $\e^{-p}$ and the problem is tractable with this promise. The exponent
$p$ depends on the particular Gaussian measure; for the Wiener measure
$p=2/3$.

We return to the promise that the smoothness $r$ is finite. Since this 
problem is intractable in the worst case, 
Wasilkowski and Wo\'zniakowski 
\cite{WW96} ask whether settling for a
stochastic assurance will break the intractability. The obvious approach is to 
approximate the infinite-dimensional integral by a $d$-dimensional integral
where $d$ may be large (or even huge) since $d$ is polynomial in $\e^{-1}$. 
Then Monte Carlo may be used since
its speed of convergence does not depend on $d$. Modulo an assumption that 
the $n$-th eigenvalue of the covariance operator of $\mu$ does not decrease
too fast the randomized query complexity is roughly $\e^{-2}$. Thus Monte 
Carlo is optimal.

\subsection{Quantum Computer}\label{sec:PIQuant}

Just as with finite dimensional integration Monte Carlo makes path integration
tractable on a classical computer and the query complexity is of 
order $\e^{-2}$. Again we ask whether we can do better on a quantum computer 
and again the short answer is yes. Under certain assumptions on the promises,
which will be made precise below, the quantum query complexity is of order
$\e^{-1}$. Thus quantum query complexity enjoys exponential speedup over the 
classical worst case query complexity and quadratic speedup over the classical
randomized query complexity. Again the latter is the same
speedup as enjoyed by Grover's search algorithm of an unstructured database.

The idea for solving path integration on a quantum computer is fairly simple
but the analysis is not so easy. So we outline
the algorithm without considering the details. We start with a
classical deterministic algorithm that uses
an average as in (\ref{eq:sum}) to 
approximate the path integral $I(f)$ with error $\e$ in the worst case.
The number of terms $N$ of the this average is an
exponential function of $\e^{-1}$.
Nevertheless, on
a quantum computer we can approximate the average, using the amplitude
amplification algorithm as we discussed in Section 4.2, with cost that
depends on $\log N$ and is, therefore, a polynomial in $\e^{-1}$ 
\cite[Sec.~6]{TW02}.

A summary of the promises and the results in \cite{TW02} follows. 
The measure $\mu$ 
is Gaussian and the eigenvalues of its covariance operator is of order
$j^{-h}$, $h>1$. For the Wiener measure occurring in many applications we have
$h=2$. The class of integrands consists of functions $f$ whose $r$-th
Fr\'echet derivatives are continuous and uniformly bounded by unity. 
In 
particular assume the integrands are at least Lipschitz. Then
\begin{itemize}
\item Path integration on a quantum computer is tractable.
\item Query complexity on a quantum computer enjoys exponential speedup over
the worst case and quadratic speedup over the classical randomized query
complexity. More precisely, the number of quantum queries is at most 
$4.22\e^{-1}$. 
\end{itemize}

Results on the qubit complexity of path integration will be given in 
Section~\ref{sec:Qubit}. Details of an algorithm for computing 
an $\e$-approximation to a path 
integral on a quantum computer may be found in \cite[Sec. 6]{TW02}.

\section{Feynman-Kac Path Integration}\label{sec:Feynman}

An important special case of a path integral is a Feynman-Kac path integral.
Assume that $X$ is the space $C$ of continuous functions and that the measure
$\mu$ is the Wiener measure $w$. Feynman-Kac path integrals occur in many
applications; see \cite{Eg93}.
For example consider the diffusion equation
\begin{eqnarray*}
&&\frac{\partial z}{\partial t}(u,t)= \frac 12 \frac{\partial^2z}{\partial t^2}
(u,t)+ V(u)z(u,t) \\
&&z(u,0)=v(u),
\end{eqnarray*}
where $u\in\reals$, $t>0$, $V$ is a potential function, and $v$ is an initial
condition function.  Under mild conditions on $v$ and $V$
the solution is given by the Feynman-Kac path integral
\begin{equation}
z(u,t)=\int_C v(x(t)+u)e^{\int_0^tV(x(s)+u)\,ds }\, w(dx).
\label{eq:F-K}
\end{equation}

The problem generalizes to the multivariate case by considering
the diffusion
equation 
\begin{eqnarray*}
&&\frac{\partial z}{\partial t}(u,t)= \frac 12 \Delta z(u,t)+ V(u)z(u,t) \\
&&z(u,0)=v(u),
\end{eqnarray*}
with $u\in\reals^d$, $t>0$, and $V,v:\reals^d\to\reals$,
the potential and the initial value function, respectively. As usual, $\Delta$
denotes the Laplacian. The solution is given by the Feynman-Kac path integral
\begin{equation*}
z(u,t)=\int_C v(x(t)+u)e^{\int_0^tV(x(s)+u)\,ds }\, w(dx),
\end{equation*}
where $C$ is the set of continuous functions $x:\reals_+\to\reals^d$ such
that $x(0)=0$.

Note that there are two kinds of dimension here. A Feynman-Kac path integral 
is infinite dimensional since we're integrating over continuous functions. 
Furthermore $u$ is a function of $d$ variables.

\subsection{Classical Computer}\label{sec:FKClass}

We begin with the case when $u$ is a scalar and then move to the case where 
$u$ is a multivariate function.
There have been a number of papers on the numerical solution of (\ref{eq:F-K});
see, for example \cite{Cam51}.

The usual attack is to solve the problem with a stochastic assurance using
randomization. For simplicity we make the promise that 
$u=1$ and $V$ is four times continuously
differentiable. Then by Chorin's algorithm \cite{Chorin}, the total cost
is of order $\e^{-2.5}$.

The first complexity analysis may be found in Plaskota et al. \cite{PWW00} 
where a new algorithm
is defined which enjoys certain optimality properties. They construct an 
algorithm which computes an $\e$-approximation at cost of order $\e^{-.25}$
and show that the worst case complexity is of the same order. Hence the 
exponent of $\e^{-1}$ is an order of magnitude smaller and with a worst case 
rather than a stochastic guarantee. However, the new algorithm requires
a numerically difficult precomputation which may limit its applicability.

We next report on multivariate Feynman-Kac path integration. 
First consider the worst case setting with the promise that $v$ and $V$ 
are $r$ times continuously differentiable with $r$ finite. 
Kwas and Li \cite{K03} proved
that the query complexity is of order $\e^{-d/r}$. Therefore in the worst 
case setting the problem suffers the curse of dimensionality.

The randomized setting was studied by Kwas \cite{K05}. He showed that 
the curse of dimensionality is broken by using Monte Carlo
using a number of queries of order $\e^{-2}$. 
The number of queries can be further improved to $\e^{-2/(1+2r/d)}$,
which is the optimal number of queries, by
a bit more complicated algorithm. The randomized algorithms
require extensive precomputing \cite{K05,K03}.

\subsection{Quantum Computer}\label{sec:FKQuant}

Multivariate multivariate Feynman-Kac path 
integration on a quantum computer was studied in \cite{K05}. 
With the promise as in the worst and randomized case, 
Kwas presents an algorithm and complexity 
analysis. He exhibits a quantum algorithm that uses a number of queries
of order $\e^{-1}$ that is based on the Monte Carlo algorithm.
He shows that the query complexity 
is of order $\e^{-1/(1+r/d)}$ and is achieved by a bit more complicated
quantum algorithm.
Just as in the randomized case the quantum algorithms require 
extensive precomputing \cite{K05,K03}.

\section{Eigenvalue Approximation}\label{sec:Eig}

Eigenvalue problems for differential operators 
arising in physics and engineering have been extensively 
studied in the literature; see, e.g. 
\cite{babuska,collatz,courant,demmel,forsythe,keller,strang,titschmarsh}.
Typically, the mathematical properties of the eigenvalues and the 
corresponding 
eigenfunctions are known and so are numerical algorithms approximating 
them on a classical computer. Nevertheless, the complexity of approximating
eigenvalues in the worst, randomized and quantum settings has only recently
been addressed for the Sturm-Liouville eigenvalue problem 
\cite{P07,PW05} (see also \cite{Bessen} for quantum lower bounds 
with a different kind of query than the one we discuss in this article). 
In some cases we have sharp complexity estimates but there are 
important questions that remain open.

Most of our discussion here concerns the complexity 
of approximating the smallest eigenvalue of a Sturm-Liouville
eigenvalue problem. We will conclude this section by briefly 
addressing quantum algorithms for other eigenvalue problems.

In the physics literature this problem is called the time-independent 
Schr\"odinger equation. The smallest eigenvalue is the energy of the
ground state. In the mathematics literature it is called the Sturm-Liouville 
eigenvalue problem.

Let $I_d=[0,1]^d$ and consider the
class of functions
$$\Q=\left\{ q:I_d\to [0,1] \bigg| \; q,\;
D_jq:=\frac{\partial q}{\partial x_j}
\in C(I_d), \; \|D_jq\|_\infty \le 1, \;\|q\|_\infty \le 1  \right\},$$
where $\|\cdot\|_\infty$ denotes the supremum norm.
For $q\in\Q$,
define $\L_q:=-\Delta + q$, where $\Delta =\sum_{j=1}^d \partial^2/
\partial x_j^2$ is the Laplacian, and consider the
eigenvalue problem
\begin{eqnarray}
&&\L_qu=\lambda u, \; x\in (0,1)^d,
\label{eq:Eval1} \\
&&u(x)\equiv 0, \; x\in \partial I_d.
\label{eq:Eval2}
\end{eqnarray}
In the variational form,
the smallest eigenvalue $\l=\l(q)$ of (\ref{eq:Eval1}, \ref{eq:Eval2}) is given
by
\begin{equation}
\l(q)=\min_{0\ne u\in H_0^1}
\frac{\int_{I_d} \sum_{j=1}^d [D_ju(x)]^2 + q(x)u^2(x)\, dx}
{\int_{I_d} u^2(x)\, dx},
\label{eq:Eval3}
\end{equation}
where $H_0^1$ is the space of all functions vanishing on the boundary of
$I_d$ having square integrable first order partial derivatives.
We consider the complexity of classical and quantum algorithms
approximating $\l(q)$ with error $\e$. 

\subsection{Classical Computer}\label{sec:EigClass}

In the worst case we discretize the differential
operator on a grid of size $h=\Theta(\e^{-1})$ and obtain a
matrix $M_\e = -\Delta_\e + B_\e$, of size proportional to 
$\e^{-d}\times \e^{-d}$. The matrix $\Delta_\e$ is the $(2d+1)$-point
finite difference discretization
of the Laplacian. The matrix  $B_\e$ is a diagonal matrix containing  
evaluations of  $q$ at the grid points. 
The smallest eigenvalue of $M_\e$ approximates $\l(q)$ with error $O(\e)$
\cite{W56,W58}.
We compute the smallest eigenvalue
of $M_\e$ using the bisection method \cite[p. 228]{demmel}.
The resulting algorithm uses a number of queries proportional to $\e^{-d}$.
It turns out that this number of queries is optimal in the worst case,
and the problem suffers from the curse of dimensionality. 

The query complexity lower bounds are obtained by reducing the eigenvalue
problem to high-dimensional integration. For this we use the perturbation 
formula \cite{P07,PW05}
\begin{equation}
\l(q)\,=\,\l(\bar q)\,+\,\int_{I_d}\left(q(x)-\bar q(x)\right)
u_{\bar q}^2(x)\,dx
\,+\, O\left(\|q-\bar q\|_{\infty}^2\right),\label{eq:333}
\end{equation}
where $q,\bar q\in\Q$ and $u_{\bar q}$ is the normalized eigenfunction
corresponding to $\l(\bar q)$.

The same formula is used for lower bounds in the randomized 
setting. Namely, the query complexity is 
$$\Omega(\e^{-2d/(d+2)}).$$
Moreover, we can use (\ref{eq:333}) to derive a randomized
algorithm.  First we approximate $q$ by a function $\bar q$ and then
approximate $\l(q)$ by 
approximating the first two terms on the right hand side of (\ref{eq:333});
see Papageorgiou and Wo\'zniakowski \cite{PW05} for $d=1$, and
Papageorgiou \cite{P07} for general $d$.
However, this algorithm uses 
$$O(\e^{-\max(2/3, d/2)}),$$
queries. So, it is optimal only when $d\le 2$. Determining 
the randomized complexity for $d>2$ and
determining if the randomized complexity
is an exponential function of $d$ are important open questions.

\subsection{Quantum Computer}\label{sec:EigQuant}

The perturbation formula (\ref{eq:333}) can be used to show that 
the quantum query complexity is
$$\Omega( \e^{-d/(d+1)} ).$$
As in the randomized case, we can use (\ref{eq:333}) to derive an algorithm
that uses $O(\e^{-d/2})$ quantum queries.
The difference between the quantum algorithm and the randomized algorithm is
that the former uses the amplitude amplification algorithm to approximate the
integral on the right hand side of (\ref{eq:333}) instead of Monte 
Carlo. The algorithm is optimal
only when $d=1$ \cite{P07,PW05}.

For general $d$ the query complexity is not known exactly, we only
have the upper bound $O(\e^{-p})$, $p\le 6$. 
The best quantum algorithm known is based on phase estimation.
In particular, we discretize the problem as in the worst case and 
apply phase estimation to the unitary matrix
$$e^{i\gamma M_\e},$$
where $\gamma$ is chosen appropriately so that the resulting phase belongs to 
$[0,2\pi)$. We use a splitting formula to approximate the necessary powers 
of the matrix exponential. 
The largest eigenvalue of $\Delta_\e$ is $O(\e^{-2})$ 
and $\|q\|_\infty \le 1$. This implies that the resulting number of queries
does not grow exponentially with $d$.

Finally, there are a number of papers providing quantum algorithms for 
eigenvalue 
approximation without carrying out a complete complexity analysis. 
Abrams and Lloyd \cite{abrams} have written an influential paper
on eigenvalue approximation of a quantum mechanical system evolving 
with a given Hamiltonian. They point out
that phase estimation, which requires the corresponding 
eigenvector as part of its initial state,
can also be used with an approximate eigenvector.
Jaksch and Papageorgiou \cite{jaksch} give a quantum algorithm 
which computes
a {\it good} approximation of the eigenvector at low cost. Their method can
be generally applied to the solution of continuous Hermitian eigenproblems on
a discrete grid. It starts with a classically obtained eigenvector for
a problem discretized on a coarse grid and constructs an approximate
eigenvector on a fine grid. 

We describe this algorithm briefly for the case $d=1$.
Suppose $N=2^k$ and $N_0=2^{k_0}$ are the number of points in the fine and
coarse grid, respectively. 
Given the eigenvector for the coarse grid $\ket{U^{(N_0)}}$, we approximate
the eigenvector for the fine grid $\ket{U^{(N)}}$ by
$$\ket{\tilde U^{(N)}}=\ket{U^{(N_0)}} 
\left( \frac{\ket{0}+\ket{1}}{\sqrt 2}\right)^{\otimes (k-k_0)}.$$
The effect of this transformation is to replicate the coordinates of 
$\ket{U^(N_0)}$ $2^{k-k_0}$ times. The resulting approximation is good enough
in the sense that the success probability of phase estimation with initial 
state $\ket{\tilde U_h}$ is greater than $\tfrac 12$.

Szkopek et al. \cite{szkopek} use the algorithm
of Jaksch and Papageorgiou in the approximation of low order eigenvalues
of a differential operator of order $2s$ in $d$ dimensions. Their
paper provides an algorithm with cost
polynomial in $\e^{-1}$ and generalizes the results of Abrams and 
Lloyd \cite{abrams}. However, \cite{szkopek} does not carry out a complexity
analysis but only considers known classical algorithms in the worst case
for comparison. 

\section{Qubit Complexity}\label{sec:Qubit}

For the foreseeable future the number of qubits will be a crucial 
computational resource. We give a general lower bound on the 
qubit complexity of continuous problems.

Recall that in (\ref{eq:qa1}) we defined a quantum algorithm as
\begin{equation*}
\ket{\psi_f}=U_TQ_fU_{T-1}Q_f\dots U_1Q_fU_0\ket{\psi_0}.
\end{equation*}
where  $\ket{\psi_0}$ and $\ket{\psi_f}$ are the initial and final state 
vectors, respectively. They are column vectors of length $2^\nu$. 
The query $Q_f$, a $2^\nu\times 2^\nu$ unitary matrix, 
depends on the
values of $f$ at $n\le 2^\nu$ deterministic points. That is
$$Q_f=Q_f(f(x_1),\dots,f(x_n)).$$
It's important to note that in the standard model the evolution is completely 
deterministic. The only probabilistic element is in the measurement of the 
final state.

For the qubit complexity (\ref{eq:qubitcomp}) 
we have the following lower bound
\begin{equation}
{\rm comp}^{\rm qubit}_{\rm std}(\e, S) =
\Omega( \log {\rm comp}^{\rm query}_{\rm clas}(2\e,S)).
\label{eq:qcstd}
\end{equation}
Here $S$ specifies the problem, see (\ref{eq:contprob}),
and ${\rm comp}^{\rm qubit}_{\rm std}(\e, S)$
is the qubit complexity in the standard
quantum setting. On the right hand side 
${\rm comp}^{\rm query}_{\rm clas}(\e,S)$ is the query complexity on 
a classical computer in the worst case setting. 
See \cite{W06} for details.

We provide an intuition about (\ref{eq:qcstd}). Assume $2^\nu$ function 
evaluations are needed to solve the problem specified by $S$ to within $\e$. 
Note that $\nu$ qubits are needed to generate the Hilbert space $\Cal H_\nu$
of size $2^\nu$ to store the evaluations.

Equation (\ref{eq:qcstd}) can be interpreted as a certain limitation of the
standard quantum setting, which considers queries using 
function evaluations at deterministic points. 
We'll show why
this is a limitation and show we can do better.

Consider multivariate integration which was discussed in 
Section~\ref{sec:Integration}.
We seek to approximate the solution of
$$S(f)=\int_{0,1]^d} f(x)\, dx.$$
Assume our promise is $f\in F_0$ where
$$F_d=\left\{ f:[0,1]^d\to\reals\; |\; \mbox{continuous and}\; |f(x)|\le 1,
x\in [0,1]^d \right\}.$$
For the moment let $d=1$.
We showed that with this promise we cannot guarantee an $\e$-approximation
on a classical computer with $\e <\tfrac 12$. That is,
$${\rm comp}^{\rm query}_{\rm clas}(\e,S)=\infty.$$
By (\ref{eq:qcstd})
$${\rm comp}^{\rm qubit}_{\rm std}(\e,S)=\infty.$$
If the qubit complexity is infinite even for $d=1$ its certainly infinite for 
general $d$. But we saw that if classical randomization (Monte Carlo) is
used
$${\rm comp}^{\rm query}_{\rm clas-ran}(\e,S)=\Theta(\e^{-2}).$$

Thus we have identified a problem which is easy to solve on a classical 
computer and is impossible to solve on a quantum computer using the
standard formulation of a quantum algorithm (\ref{eq:qa1}).
Our example motivates extending the notion of quantum algorithm
by permitting {\it randomized queries}. 
The quantum setting with randomized queries was introduced by 
Wo\'zniakowski~\cite{W06}.
The idea of using randomized queries is not entirely new. Shor's algorithm
\cite{shor} uses a special kind of randomized query, namely,
$$Q_x\ket{x}=\ket{jx \;\mbox{mod} N},$$
with $j=0,\dots,N-1$ and a random $x$ from the set $\{ 2,3,\dots,N-1\}$.

In our case, the randomization
affects only the selection of sample points and the number of queries.
It occurs prior to the implementation of the queries and the
execution of the quantum algorithm.
In this extended setting, we define a quantum algorithm as
\begin{equation}
\ket{\psi_{f,\omega}}=U_{T_\omega}Q_{f,\omega}U_{T_\omega -1}Q_{f,\omega}
\dots U_1Q_{f,\omega} U_0\ket{\psi_0},
\label{eq:rqa}
\end{equation}
where $\omega$ is a random variable and
$$Q_{f,\omega}=Q_{f,\omega}(f(x_{1,\omega}),\dots,f(x_{n,\omega})),$$
and the $x_{j,\omega}$ are random points. The number of queries $T_\omega$
and the points $x_{j,\omega}$ are chosen at random initially and then remain 
fixed for the remainder of the computation. Note that (\ref{eq:rqa}) is
identical to (\ref{eq:qa1}) except for the introduction of randomization.
Randomized queries require that we modify the criterion (\ref{eq:wperr})
by which we measure the error of an algorithm. One possibility is
to consider the expected error and another possibility is to consider 
the probabilistic error with respect to the random variable $\omega$. 
Both cases have been considered in the literature \cite{W06} 
but we will avoid the details here because they are rather technical.

A test of the new setting is whether it buys us anything. We compare the qubit
complexity of the standard and randomized settings for integration and
path integration.

\subsection{Integration}\label{sec:QubitInt}
We make the same promise as above, namely, $f\in F_d$.
\begin{itemize}
\item Quantum setting with deterministic queries. 
We remind the reader that
\begin{eqnarray*}
&&{\rm comp}^{\rm query}_{\rm std}(\e)=\infty\\
&&{\rm comp}^{\rm qubit}_{\rm std}(\e)=\infty.
\end{eqnarray*}
\item Quantum setting with randomized queries.
Then \cite{W06}
\begin{eqnarray*}
&&{\rm comp}^{\rm query}_{\rm ran}(\e)=\Theta(\e^{-1})\\
&&{\rm comp}^{\rm qubit}_{\rm ran}(\e)=\Theta(\log\e^{-1}).
\end{eqnarray*}
Therefore, there is infinite improvement in the randomized quantum setting over
the standard quantum setting.
\end{itemize}

\subsection{Path integration}\label{sec:QubitPI}

\begin{itemize}
\item Quantum setting with deterministic queries.
With an appropriate promise it was shown \cite{TW02} that
\begin{eqnarray*}
&&{\rm comp}^{\rm query}_{\rm std}(\e)=\Theta(\e^{-1})\\
&&{\rm comp}^{\rm qubit}_{\rm std}(\e)=\Theta(\e^{-2}\log\e^{-1}).
\end{eqnarray*}
Thus, modulo the $\log$ factor, the qubit complexity of path integration is a 
second degree polynomial in $\e^{-1}$. That seems pretty good but we probably 
won't have enough qubits for a long time to do new science, especially with error correction.
\item Quantum setting with randomized queries.
Then \cite{W06}
\begin{eqnarray*}
&&{\rm comp}^{\rm query}_{\rm ran}(\e)=\Theta(\e^{-1})\\
&&{\rm comp}^{\rm qubit}_{\rm ran}(\e)=\Theta(\log\e^{-1}).
\end{eqnarray*}
Thus there is an exponential improvement in the randomized quantum setting
over the standard quantum setting.
\end{itemize}

As the analogue of (\ref{eq:qcstd}) we have the following lower bound on the 
randomized qubit complexity \cite{W06}
\begin{equation}
{\rm comp}^{\rm qubit}_{\rm ran}(\e, S) = 
\Omega( \log {\rm comp}^{\rm query}_{\rm clas-ran}(\e,S)),
\label{eq:qcran}
\end{equation}
where ${\rm comp}^{\rm query}_{\rm clas-ran}(\e,S)$
is the query complexity on a classical computer in the randomized setting.

\section{Approximation}\label{sec:Approx}

Approximating functions of $d$ variables is a fundamental 
and generally hard problem. Typically, the literature considers
the approximation of functions that belong to the Sobolev
space $W_p^r([0,1]^d)$ in the norm of $L_q([0,1]^d)$. 
The condition $r/d > 1/p$ ensures that 
functions in $W_p^r([0,1]^d)$ are continuous, which is
necessary for function values to be well defined.
Thus, when $p=\infty$ the dimension $d$ can be arbitrarily large while the 
smoothness $r$ can be fixed, which cannot happen when $p<\infty$.

For $p=\infty$ 
the problem suffers the curse of dimensionality
in the worst and the randomized classical cases \cite{novak,TWW88}.
Recently, Heinrich \cite{H04b} showed that quantum computers
do not offer any advantage relative to classical computers since
the problem remains intractable in the quantum setting.

For different values of the parameters $p,q,r,d$ the classical and quantum 
complexities are also known \cite{H04b,novak,TWW88}. In some cases
quantum computers can provide a roughly quadratic speedup over classical 
computers, but there are also cases where the classical and quantum
complexities coincide. The table below summarizes the order of the 
query complexity (up to polylog factors) of approximation 
in the worst case, randomized and quantum setting, and 
is based on a similar table in \cite{H04b}
describing error bounds.
\begin{center}
\begin{tabular}{|l|l|l|l|}
\hline
             & Worst case          & Randomized             & Quantum \\
\hline 
\hline
$1\le p<q\le\infty$, & $\e^{-dpq/(rpq-d(q-p))}$  & $\e^{-dpq/(rpq-d(q-p))}$ &  $\e^{-d/r}$  \\
$r/d\ge2/p-2/q$ &   &    &  \\
  &   &  &   \\
$1\le p<q\le\infty$, & $\e^{-dpq/(rpq-d(q-p))}$  & $\e^{-dpq/(rpq-d(q-p))}$  & $\e^{-dpq/(2rpq-2d(q-p))}$  \\
$r/d<2/p-2/q$  &   &    &  \\
   &    &   &    \\
$1\le q\le p\le\infty$ & $\e^{-d/r}$  & $\e^{-d/r}$  & $\e^{-d/r}$ \\
\hline
\end{tabular}
\end{center}

\section{Elliptic Partial Differential Equations}\label{sec:PDE}

Elliptic partial differential equations have many important applications
and have been extensively studied in 
the literature, see \cite{Wer91} and the references therein. 
A simple
example is the Poisson equation, for which we want to find a function 
$u:\bar \Omega\to\reals$, 
that satisfies
\begin{eqnarray*}
&&-\Delta u(x) = f(x),\quad x\in \Omega\\
&&\; u(x)=0, \quad x\in \partial\Omega,
\end{eqnarray*}
where $\Omega \subset \reals^d$,

More generally we consider elliptic partial differential 
equations of order $2m$ on a
smooth bounded domain $\Omega\subset \reals^d$ with smooth
coefficients and homogeneous boundary 
conditions with the right hand side belonging to $C^r(\Omega)$ and 
the error measured in the $L_{\infty}$ norm; see \cite{H06b} for details.

In the worst case the complexity is proportional to $\e^{-d/r}$
\cite{Wer91} and the problem is intractable. 
The randomized complexity of this problem was only recently 
studied along with the quantum complexity by Heinrich \cite{H06a,H06b}.
In particular, the randomized query complexity (up to polylog factors)
is proportional to
$$\e^{-\max\{ d/(r+2m), \, 2d/(2r+d)\}},$$
and the quantum query complexity is proportional to
$$\e^{-\max\{ d/(r+2m), \, d/(r+d)\}}.$$
Thus the quantum setting may provide a polynomial speedup over the
classical randomized setting but not always.
Moreover, 
for fixed $m$ and $r$ and for $d>4m$ the problem 
is intractable in all three settings.

\section{Ordinary Differential Equations}\label{sec:IVP}

In this section we consider the solution of a system of ordinary differential
equations with initial conditions
$$z^\prime(t)=f(z(t)),\quad t\in[a,b],\quad z(a)=\eta,$$
where $f:\reals^d\to\reals^d$, $z:[a,b]\to\reals^d$ and $\eta\in\reals^d$
with $f(\eta)\ne 0$. For the right hand side function 
$f=[f_1,\dots,f_d]$, where $f_j:\reals^d\to\reals$, 
we assume that the $f_j$ belong
to the H\"older class $F_d^{r,\a}$, $r+\a\geq 1$. 
We seek to compute a bounded function
on the interval $[a,b]$ that approximates the solution~$z$.

Kacewicz \cite{Ka84} studied the classical worst case complexity of 
this problem and
found it to be proportional to $\e^{-1/(r+\a)}$. Recently he also studied
the classical randomized and quantum complexity of the problem and derived
algorithms that yield upper bounds that from the lower bounds by only
an arbitrarily small positive parameter in the exponent \cite{Ka06}. 
The resulting randomized and quantum complexity bounds (up to polylog factors)
are
$$O( \e^{-1/(r+\a+1/2-\c)} )$$
and
$$O( \e^{-1/(r+\a+1-\c)} ),$$
where $\c\in (0,1)$ is arbitrarily small, respectively.
Observe that 
the randomized and quantum complexities (up to polylog factors)
satisfy
$$\Omega(\e^{-1/(r+\a+1/2)})$$
and
$$\Omega(\e^{-1/(r+\a+1)}),$$
respectively. 
Even more recently, Heinrich and Milla \cite{HM07} showed that the upper 
bound for the randomized complexity holds with $\c= 0$, thereby establishing
tight upper and lower randomized complexity bounds, up to polylog factors.

Once more, the quantum setting provides a polynomial speedup over the
classical setting.

\section{Gradient Estimation}\label{sec:Gradient}

Approximating the gradient of a function $f:\reals^d \to \reals$ 
with accuracy $\e$ requires a minimum of
$d+1$ function evaluations on a classical computer. Jordan \cite{Jo}
shows how this 
can be done using a single query on a quantum computer. 

We present Jordan's algorithm for the special case where
the function is a plane passing through the origin, i.e.,
$f(x_1,\dots,x_d)=\sum_{j=1}^d a_j x_j$, and is uniformly
bounded by $1$. 
Then $\nabla f = (a_1,\dots,a_d)^T$.
Using a single query and {\it phase kickback} we obtain the state
$$
\frac 1{\sqrt{N^d}}\sum_{j_1=0}^{N-1} \cdots \sum_{j_d=0}^{N-1}
e^{2\pi i f(j_1,\dots,j_d) } \ket{j_1}\cdots \ket{j_d},
$$  
where $N$ is a power of $2$.
Equivalently, we have
$$
\frac 1{\sqrt{N^d}}\sum_{j_1=0}^{N-1} \cdots \sum_{j_d=0}^{N-1}
e^{2\pi i ( a_1 j_1 +\cdots + a_d j_d )} 
\ket{j_1}\cdots \ket{j_d} .$$
This is equal to the state
$$\frac 1{\sqrt{N}} \sum_{j_1=0}^{N-1} e^{2\pi i a_1 j_1} \ket{j_1}
\dots \frac 1{\sqrt{N}} \sum_{j_d=0}^{N-1} e^{2\pi i a_d j_d} \ket{j_d}.$$
We apply the Fourier transform to each of the $d$ 
registers and then measure 
each register in the computational basis to obtain $m_1,\dots,m_d$.
If $a_j$ can be represented with finitely many bits and
$N$ is sufficiently large 
then $m_j/ N=a_j$, $j=1,\dots,d$. 

For functions with second order partial derivatives
not identically equal to zero the analysis is more complicated and we refer
the reader to \cite{Jo} for the details.

\section{Simulation of Quantum Systems on Quantum Computers}\label{sec:Sim}

So far this article has been devoted to work on algorithms and complexity of
problems where the query and qubit complexities are known or
have been studied. In a number of cases, the
classical complexity of these problems is also known
and we know the quantum computing speedup.

The notion that quantum systems could be simulated more efficiently by
quantum computers than by classical computers was first mentioned by 
Manin \cite{Ma80}, see also \cite{Ma99}, and discussed 
thoroughly by Feynman \cite{Fey82}. 

There is a large and varied literature on simulation of quantum systems on 
quantum computers. The focus in these papers is typically on the cost
of particular quantum and classical algorithms without complexity
analysis and therefore without speedup results.
To give the reader a taste of this area we list some sample papers:
\begin{itemize}
\item Berry et al. \cite{Be07} present an efficient quantum algorithm for
simulating the evolution of a sparse Hamiltonian.
\item Dawson, Eisert and Osborne \cite{Da07} introduce a unified formulation
of variational methods for simulating ground state properties of quantum
many-body systems.
\item Morita and Nishimori \cite{Mo07} derive convergence conditions
for the quantum annealing algorithm.
\item Brown, Clark, Chuang \cite{Br06} establish limits of quantum simulation
when applied to specific problems.
\item Chen, Yepez and Cory \cite{Che06} report on the simulation of
Burgers equation as a type-II quantum computation.
\item Paredes, Verstraete and Cirac \cite{Pa05} present an algorithm
that exploits quantum parallelism to simulate randomness.
\item Somma et al. \cite{So03} discuss what type of physical problems can be 
efficiently simulated on a quantum computer which cannot be simulated
on a Turing machine.
\item Yepez \cite{Ye02} presents an efficient algorithm for the many-body 
three-dimensional Dirac equation.
\item Nielsen and Chuang \cite{NC} discuss simulation of a variety
of quantum systems.
\item Sornborger and Stewart \cite{Sor99}
develop higher order methods for simulations.
\item Boghosian and Taylor \cite{Bo98} present algorithms for efficiently
simulating quantum mechanical systems.
\item Zalka \cite{zalka} shows that the time evolution of the wave 
function of a quantum mechanical
many particle system can be simulated efficiently.
\item Abrams and Lloyd \cite{AL97} provide fast algorithms for simulating 
many-body Fermi systems.
\item Wisner \cite{Wi96} provides two quantum many-body problems 
whose solution is intractable on
a classical computer.
\end{itemize}

\section{Future Directions}\label{sec:FutDir}

The reason there is so much interest in quantum computers is 
to solve important problems fast. The belief is
that we will be able to solve scientific problems, and in particular
quantum mechanical systems, which cannot be solved on a classical
computer. That is, that quantum computation will lead to new science.

Research consists of two major parts. The first is identification
of important scientific problems with substantial 
speedup. The second is the construction of machines with sufficient number of 
qubits and long enough decoherence times to solve the problems identified
in the first part. Abrams and Lloyd \cite{abrams} have
argued that with 50 to 100 qubits we can solve interesting classically
intractable problems from atomic physics. 
Of course this does not include qubits
needed for fault tolerant computation. 

There are numerous important open questions.
We will limit ourselves here to some open questions regarding 
the problems discussed in this article.
\begin{enumerate}
\item In Section~\ref{sec:PathInt} 
we reported big wins for the qubit complexity for
integration and path integration. Are there big wins for other problems?
\item Are there problems for which we get big wins for query complexity
using randomized queries?
\item Are there tradeoffs between the query complexity and the qubit 
complexity?
\item What are the classical and quantum complexities of approximating 
the solution of the Schr\"odinger equation for a many-particle system?  
How do they it depend on the number of particles?
What is the quantum speedup?
\end{enumerate}

\section*{Acknowledgements}
J. F. Traub is an external professor at the Santa Fe Institute.
The research was supported in part
by the National Science Foundation.

We are grateful to Erich Novak, University of Jena, and Henryk Wo\'zniakowski,
Columbia University and University of Warsaw, for their very helpful comments.
We thank Jason Petras, Columbia University, for checking the complexity 
estimates appearing in the tables.


\begin{thebibliography} {99}

\bibitem{AL97}
Abrams, D. S. and Lloyd, S. (1997),
Simulation of Many-Body Fermi Systems on a Universal Quantum Computer,
{\it Phys. Rev. Lett.}, 79(13), 2586--2589. Also
Also http://arXiv.org/quant-ph/9703054.

\bibitem{abrams}
Abrams, D. S. and Lloyd, S. (1999),
Quantum Algorithm Providing Exponential Speed Increase for 
Finding Eigenvalues and Eigenvectors,
{\it Phys. Rev. Lett.}, 83, 5162--5165.

\bibitem{AW99}
Abrams, D. S. and Williams, C. P. (1999),
Fast quantum algorithms for numerical integrals and
stochastic processes, http://arXiv.org/quant-ph/9908083.

\bibitem{Bakh77}
Bakhvalov, N. S. (1977),
Numerical Methods, Mir Publishers, Moscow.

\bibitem{babuska}
Babuska, I. and Osborn, J. (1991), 
Eigenvalue Problems, in Handbook of Numerical Analysis, Vol. II, 
P. G. Ciarlet and J. L. Lions, eds., North-Holland, Amsterdam, 641--787.

\bibitem{beals}
Beals, R., Buhrman, H., Cleve, R., Mosca, R. and de Wolf, R. (1998), 
Quantum lower bounds by polynomials, Proceedings FOCS'98, 352--361.
Also http://arXiv.org/quant-ph/9802049.

\bibitem{bennet}
Bennett, C. H., Bernstein, E., Brassard, G. and Vazirani, U. (1997)
Strengths and weaknesses of quantum computing,
{\it SIAM J. Computing}, 26(5), 1510--1523.

\bibitem{bernstein}
Bernstein, E., and Vazirani, U. (1997)
Quantum complexity theory,
{\it SIAM J. Computing}, 26(5), 1411--1473.

\bibitem{Be07}
Berry, D. W., Ahokas, G., Cleve, R., Sanders, B. C. (2007),
Efficient quantum algorithms for simulating sparse Hamiltonians,
{\it Communications in Mathematical Physics}, 270(2), 359--371.
Also http://arXiv.org/quant-ph/0508139

\bibitem{Bessen}
Bessen, A. J. (2007), 
On the complexity of classical and quantum algorithms for
numerical problems in quantum mechanics, Ph.D. Thesis, Department of
Computer Science, Columbia University.

\bibitem{Bo98}
Boghosian, B. M. and Taylor, W. (1998),
Simulating quantum mechanics on a quantum computer,
Proceedings of the fourth workshop on Physics and computation,
Boston, Massachusetts, 30--42.   
Also http://arXiv.org/quant-ph/9701019

\bibitem{BHMT}
Brassard, G., Hoyer, P., Mosca, M., and Tapp, A. (2002),
Quantum Amplitude Amplification and Estimation,
in {\it Contemporary Mathematics}, Vol. 305, Am. Math. Soc., 53--74. 
Also  http://arXiv.org/quant-ph/0005055.

\bibitem{Br06}
Brown, K. R., Clark, R. J. and Chuang, I. L. (2006),
Limitations of Quantum Simulation Examined by Simulating a Pairing Hamiltonian
using Magnetic Resonance, 
{\it Phys. Rev. Lett.}, 97(5), 050504.
Also  http://arXiv.org/quant-ph/0601021.

\bibitem{Cam51}
Cameron, R. H. (1951),
A Simpson's rule for the numerical evaluation of Wiener's 
integrals in function space,
{\it Duke Math. J.}, 8, 111--130.

\bibitem{Che06}
Chen, Z., Yepez, J. and Cory, D. G. (2006),
Simulation of the Burgers equation by NMR quantum information processing,
{\it Phys. Rev. A}, 74, 042321.
Also  http://arXiv.org/quant-ph/0410198.

\bibitem{Chorin}
Chorin, A. J. (1973),
Accurate evaluation of Wiener integrals,
{\it Math. Comp.}, 27, 1--15.

\bibitem{cleve}
Cleve, R., Ekert, A., Macchiavello, C. and Mosca, M. (1996),
Quantum Algorithms Revisited,
{\it Proc. R. Soc. Lond. A.}, 454(1969), 339-354.

\bibitem{collatz}
Collatz, L. (1960),
The Numerical Treatment of Differential Equations,
Springer-Verlag, Berlin.

\bibitem{courant}
Courant, C. and Hilbert, D. (1989),
Methods of Mathematical Physics, 
Vol. I, Wiley Classics Library,
Wiley-Interscience, New York.

\bibitem{Cur00}
Curbera, F. (2000),
Delayed curse of dimension for Gaussian integration,
{\it J. Complexity}, 16(2), 474--506.

\bibitem{Da07}
Dawson, C. M., Eisert, J. and Osborne, T. J. (2007),
Unifying variational methods for simulating quantum many-body systems,
http://arxiv.org/abs/0705.3456v1.

\bibitem{demmel}
Demmel, J. W. (1997),
Applied Numerical Linear Algebra, SIAM, Philadelphia.

\bibitem{Eg93}
Egorov, A. D., Sobolevsky, P. I. and Yanovich, L. A. (1993),
Functional Integrals: Approximate Evaluation and Applications,
Kluwer Academic Publishers, Dordrecht.

\bibitem{Fey82}
Feynman, R. P. (1982),
Simulating physics with computers,
{\it Int. J. Theor. Phys.}, 21:476.

\bibitem{forsythe}
Forsythe, G. E., and Wasow, W. R. (2004),
Finite-Difference Methods for Partial Differential Equations,
Dover, New York.


\bibitem{Grover}
Grover, L. (1997),
Quantum mechanics helps in searching for a needle
in a haystack, {\it Phys. Rev. Lett.}, 79(2), 325--328.
Also http://arXiv.org/quant-ph/9706033.

\bibitem{heinrich}
Heinrich, S. (2002), 
Quantum Summation with an Application to Integration, 
{\it J. Complexity}, 18(1), 1--50. 
Also http://arXiv.org/quant-ph/0105116.

\bibitem{H03b}
Heinrich, S. (2003),
From Monte Carlo to Quantum Computation,
Proceedings of the 3rd IMACS Seminar on Monte Carlo Methods 
MCM2001, Salzburg, {\it Special Issue of Mathematics and Computers in  
Simulation}, K. Entacher, W. Ch. Schmid, A. Uhl, eds., 62, 219--230.

\bibitem{H03}
Heinrich, S. (2003), 
Quantum integration in Sobolev spaces, 
{\it J. Complexity}, 19, 19--42. 


\bibitem{H04b}
Heinrich, S. (2004),
Quantum Approximation II. Sobolev Embeddings,
{\it J. Complexity}, 20, 27--45. Also http://arXiv.org/quant-ph/0305031.

\bibitem{H06a}
Heinrich, S. (2006),
The randomized complexity of elliptic PDE,
{\it J. Complexity}, 22(2), 220--249.

\bibitem{H06b}
Heinrich, S. (2006),
The quantum query complexity of elliptic PDE,
{\it J. Complexity}, 22(5), 691--725.

\bibitem{HKW}
Heinrich, S., Kwas, M. and Wo\'zniakowski, H. (2004),
Quantum Boolean Summation with Repetitions in the Worst-Average Setting,
in Monte Carlo and Quasi-Monte Carlo Methods 2002, H. Niederreiter, ed.,
Spinger-Verlag, 27--49.   

\bibitem{HM07}
Heinrich, S. and Milla, B. (2007),
The randomized complexity of initial value problems,
Talk presented at First Joint Intermational Meeting between the American
Mathematical Society and the Polish Matehmatical Society, Warsaw, Poland.

\bibitem{HN02}
Heinrich, S. and Novak, E. (2002),
Optimal summation by deterministic, randomized and quantum algorithms,
in Monte Carlo and Quasi-Monte Carlo Methods 2000, 
K.-T. Fang, F. J. Hickernell and H. Niederreiter eds., Springer-Verlag, Berlin.

\bibitem{jaksch}
Jaksch, P. and Papageorgiou, A. (2003),
Eigenvector approximation leading to exponential speedup of quantum
eigenvalue calculation, 
{\it Phys. Rev. Lett.}, 91, 257902. 
Also http://arXiv.org/quant-ph/0308016.

\bibitem{Jo}
Jordan, S. P. (2005),
Fast Quantum Algorithm for Numerical Gradient Estimation,
{Phys. Rev. Lett.}, 95, 050501. 
Also http://arXiv.org/quant-ph/0405146.

\bibitem{Ka84}
Kacewicz, B. Z. (1984),
How to increase the order to get minimal-error algorithms
for systems of ODEs,
{\it Numer. Math.}, 45, 93-104.

\bibitem{Ka06}
Kacewicz, B. Z. (2006),
Almost optimal solution of initial-value problems by randomized and
quantum algorithms.
{\it J. Complexity}, 22(5), 676--690.

\bibitem{keller}
Keller, H. B. (1968),
Numerical methods for two-point boundary-value problems, 
Blaisdell Pub. Co., Waltham, MA.

\bibitem{Knuth}
Knuth, D. E. (1997),
The Art of Computer Programming Volume:2 Seminumerical Algorithms,
Addison-Wesley Professional, 3rd edition., Cambridge, MA.

\bibitem{K05}
Kwas, M. (2005),
Quantum algorithms and complexity for certain continuous and
related discrete problems,
Ph.D. thesis, Department of Computer Science, Columbia University.

\bibitem{K03}
Kwas, M. and Li, Y. (2003),
Worst case complexity of multivariate Feynman-Kac path integration,
{\it J. Complexity}, 19, 730--743.


\bibitem{Ma80}
Manin, Y. (1980),
Computable and Uncomputable, Sovetskoye Radio, Moscow (in Russian).

\bibitem{Ma99}
Manin, Y. I. (1999),
Classical computing, quantum computing, and Shor's
factoring algorithm, http://arXiv.org/quant-ph/9903008.

\bibitem{Mo07}
Morita, S. and Nishimori, H. (2007),
Convergence of Quantum Annealing with Real-Time Schr\"odinger Dynamics,
{\it Journal of the Physical Society of Japan}, 76(6), 064002.
Also http://arXiv.org/quant-ph/0702252.

\bibitem{nayak}
Nayak, A. and Wu, F. (1999), 
The quantum query complexity of approximating the median
and related statistics, 
Proc. STOC 1999, 384--393. Also
http://arXiv.org/quant-ph/9804066.

\bibitem{NC}
Nielsen, M.A. and Chuang, I.L. (2000), Quantum Computation and
Quantum Information, Cambridge University Press, Cambridge, UK.

\bibitem{novak}
Novak, E. (1988),
Deterministic and Stochastic Error Bounds in Numerical Analysis,
Lecture Notes in Mathematics 1349, Springer-Verlag, Berlin.

\bibitem{N01}
Novak, E. (2001), 
Quantum complexity of integration,
{\it J. Complexity}, 17, 2--16.
Also http://arXiv.org/quant-ph/0008124. 

\bibitem{Or01}
Ortiz, G., Gubernatis, J. E., Knill, E. and Laflamme, R. (2001)
Quantum algorithms for fermionic simulations,
{\it Phys. Rev. A}, 64(2), 022319.
Also http://arXiv.org/cond-mat/0012334

\bibitem{P04}
Papageorgiou, A. (2004),
Average case quantum lower bounds for computing the boolean mean,
{\it J. Complexity}, 20(5), 713--731.

\bibitem{P07}
Papageorgiou, A. (2007),
On the complexity of the multivariate
Sturm-Liouville eigenvalue problem,
{\it J. Complexity} (to appear).

\bibitem{PT05}
Papageorgiou, A. and Traub, J. F. (2005),
Qubit complexity of continuous problems, 
http://arXiv.org/quant-ph/0512082.

\bibitem{PW05}
Papageorgiou, A. and Wo\'zniakowski, H. (2005),
Classical and Quantum Complexity of the Sturm-Liouville
Eigenvalue Problem, 
{\it Quantum Information Processing}, 4(2), 87--127.
Also http://arXiv.org/quant-ph/0502054.

\bibitem{Pa05}
Paredes, B., Verstraete, F. and Cirac, J. I. (2005),
Exploiting Quantum Parallelism to Simulate Quantum Random Many-Body Systems,
{\it Phys. Rev. Lett.}, 95, 140501. 
Also http://arXiv.org/cond-mat/0505288.

\bibitem{Plask96}
Plaskota, L. (1996),
Noisy Information and Computational Complexity,
Cambridge University Press, Cambridge, UK.

\bibitem{PWW00}
Plaskota, L., Wasilkowski, G. W. and Wo\'zniakowski, H. (2000),
A new algorithm and worst case complexity for Feynman-Kac path integration,
{\it J. Comp. Phys.}, 164(2), 335--353.

\bibitem{Ritter00}
Ritter, K. (2000),
Average-Case Analysis of Numerical Problems,
Lecture Notes in Mathematics, 1733, Springer, Berlin.

\bibitem{shor}
Shor, P. W. (1997),
Polynomial-time algorithms for prime factorization and
discrete logarithm on a quantum computer, 
{\it SIAM J. Comput.}, 26(5), 1484--1509.


\bibitem{So03}
Somma, R., Ortiz, G., Knill, E., Gubernatis (2003),
Quantum Simulations of Physics Problems,
in Proc. SPIE 2003, Volume 5105, Quantum Information and Computation,
Andrew R. Pirich, Howard E. Brant eds., 96--103. Also 
http://arXiv.org/quant-ph/0304063.

\bibitem{Sor99}
Sornborger, A. T. and Stewart, E. D. (1999),
Higher Order Methods for Simulations on Quantum Computers,
{\it Phys. Rev. A}, 60(3), 1956--1965.
Also http://arXiv.org/quant-ph/9903055.

\bibitem{strang}
Strang, G. and Fix, G. J. (1973),
An Analysis of the Finite Element Method, Prentice-Hall,
Englewood Cliffs, NJ.

\bibitem{szkopek}
Szkopek, T., Roychowdhury, V., Yablonovitch, E. and Abrams, D. S. (2005),
Egenvalue estimation of differential operators with a quantum algorithm,
{\it Phys. Rev. A}, 72, 062318.

\bibitem{titschmarsh}
Titschmarsh, E. C. (1958),
Eigenfunction Expansions Associated with Second-Order
Differential Equations, Part B, Oxford University Press, Oxford, UK.

\bibitem{T99}
Traub, J. F. (1999),
A continuous model of computation,
{\it Physics Today}, May, 39-43.

\bibitem{TWW88}
Traub, J. F., Wasilkowski, G. W. and Wo\'zniakowski, H. (1988),
Information-Based Complexity, Academic Press, New York.

\bibitem{TW98}
Traub, J. F., and Werschulz, A. G. (1998),
Complexity and Information,
Cambridge University Press, Cambridge, UK. 

\bibitem{TW80}
Traub, J. F. and Wo\'zniakowski, H. (1980),
A general theory of optimal algorithms,
ACM Monograph Series, Academic Press, New York.

\bibitem{TW92}
Traub, J. F. and Wo\'zniakowski, H. (1992),
The Monte Carlo algorithm with a pseudorandom generator,
{\it Math. Comp.}, 58(197), 323--339.

\bibitem{TW02}
Traub, J. F. and Wo\'zniakowski, H. (2002),
Path integration on a quantum computer,
{\it Quantum Information Processing}, 1(5), 365--388, 2002. 
Also  http://arXiv.org/quant-ph/0109113.

\bibitem{WW96}
Wasilkowski, G. W. and Wo\'zniakowski, H. (1996),
On tractability of path integration, 
{\it J. Math. Phys.}, 37(4), 2071--2088.

\bibitem{W56}
Weinberger, H. F. (1956),
{\em Upper and Lower Bounds for Eigenvalues by Finite Difference Methods,}
Communications on Pure and Applied Mathematics, IX, 613--623.

\bibitem{W58}
Weinberger, H. F. (1958),
{\em Lower Bounds for Higher Eigenvalues by Finite Difference Methods,}
Pacific Journal of Mathematics, 8(2), 339--368.

\bibitem{Wer91}
Werschulz, A. G. (1991),
The Computational Complexity of Differential and Integral Equations,
Oxford University Press, Oxford.


\bibitem{Wi96}
Wisner, S. (1996),
Simulations of Many-Body Quantum Systems by a Quantum Computer,
http://arXiv.org/quant-ph/96

\bibitem{W06}
Wo\'zniakowski, H. (2006),
The Quantum Setting with Randomized Queries for Continuous Problems,
{\it Quantum Information Processing}, 5(2), 83--130.

\bibitem{Ye02}
Yepez, J. (2002),
An efficient and accurate quantum algorithm for the Dirac equation,
http://arXiv.org/quant-ph/0210093.

\bibitem{zalka}
Zalka, C. (1998),
Simulating quantum systems on a quantum computer, 
{\it Proc. Royal Soc. Lond. A}, 454(1969), 313--322.
Also  http://arXiv.org/quant-ph/9603026.

\end{thebibliography}

\end{document}